\documentclass[3p,times,twocolumn]{elsarticle}

\usepackage{ecrc}
\usepackage{amsmath,amssymb,latexsym,color}

\newcommand{\mpl}{m_p}
\volume{00}

\firstpage{1}

\journalname{Nuclear Physics B Proceedings Supplement}

\runauth{J. F. Rodr\'{\i}guez and Y. Rodr\'{\i}guez}

\jid{nuphbp}

\jnltitlelogo{Nuclear Physics B Proceedings Supplement}

\usepackage{amssymb}
\graphicspath{{figures/}} % Location of the graphics files
\usepackage[figuresright]{rotating}

\begin{document}

\begin{frontmatter}



\dochead{PI/UAN-2015-582FT}

\title{Analysis of Vector-Inflation Models Using Dynamical Systems}

\author[label1]{Jos\'e F. Rodr\'{\i}guez}
\ead{jose.rodriguez2@correo.uis.edu.co}
\author[label1,label2]{Yeinzon Rodr\'{\i}guez}
\ead{yeinzon.rodriguez@uan.edu.co}
\address[label1]{Escuela de F\'isica, Universidad Industrial de Santander, Ciudad Universitaria, Bucaramanga 680002, Colombia}
\address[label2]{Centro de Investigaciones en Ciencias B\'asicas y Aplicadas, Universidad Antonio Nari\~no, Cra 3 Este \# 47A -15, Bogot\'a D.C. 110231, Colombia}

\begin{abstract}
We analyze two possible vector-field models using the techniques of dynamical systems. The first model involves a U(1)-vector field and the second a triad of SU(2)-vector fields. Both models include a gauge-fixing term and a power-law potential. A dynamical system is formulated and it is found that one of the critical points, for each model, corresponds to inflation, the origin of these critical points being the respective gauge-fixing terms. The conditions for the existence of an inflationary era which lasts for at least 60 efolds are studied.
\end{abstract}


\end{frontmatter}

% \linenumbers

\section{Introduction}
\label{intro}
Scalar-field inflation provides a satisfactory solution to the flatness,
horizon and unwanted relics problems.
Furthermore, the quantum primordial fluctuations together with the expansion of the
Universe generate the observed large-scale structure. 
However, the observations of the Planck mission hint towards a privileged direction in the cosmic microwave background fluctuations \cite{Ade:2013nlj}. One plausible scenario to explain such a privileged direction
consists in adding vector fields to the inflationary mechanism.
Due to the nonlinearity of the field equations, an analytical solution 
may be difficult, if not impossible.
In this work, we study two inflation model candidates which involve vector fields together with a gauge-fixing term.
The analysis is done by using numerical methods and specially the techniques
of dynamical systems.
We look for the attractors of the models and other qualitative behaviours.
The main purpose of this proceedings contribution is to find whether the attractors of the system correspond
to an inflationary period.
%We also examine whether the level of anisotropic expansion is in agreement with the observational data.
\section{Spacetime Geometry}
	To analyze the dynamics, we assume a homogeneous but anisotropic spacetime. We choose a Bianchi-Type-I metric expressed in the following way:
		\begin{equation}
			g_{tt}=1, \quad g_{xx}=g_{yy}=-e^{2(\alpha+\sigma)},\quad g_{zz}=-e^{2\alpha-4\sigma},
               		\label{eqn:bianchy_typeI}
       		\end{equation}
where $\alpha$ and $\sigma$ are functions of time. The dynamics of the system can be expressed in terms of the following kinematical quantities: the global Hubble parameter which measures the volume expansion,
%\begin{equation}
%    H\equiv\dot{\alpha},
%    \label{eqn:globalhubble}
%\end{equation}
%        and the cosmic shear which measures the anisotropic expansion:
%\begin{equation}
%            \Sigma\equiv\frac{H_{\text{xy}}-H}{H}=\frac{\dot{\sigma}}{\dot{\alpha}},
%            \label{eqn:shear}
%\end{equation}
%where $H_{xy}$ is the Hubble parameter on spacelike hypersuperfaces with constant coordinate $z$.  The dot above some quantities means derivation with respect to the cosmic time.
$H\equiv\dot{\alpha}$, and the cosmic shear which measures the anisotropic expansion,  $\Sigma\equiv\frac{\dot{\sigma}}{\dot{\alpha}}$. The dot above some quantities means derivation with respect to the cosmic time.
%----------------------------------------------------------------------------------------
%	U(1) Model
%----------------------------------------------------------------------------------------

\section{U(1)-Vector-Field Model}
The matter lagrangian density for the U(1)-vector field $A_{\mu}$ is \cite{Cembranos:2012kk}: 
\begin{equation}
\mathcal{L}_M= -\frac{1}{4}F_{\mu\nu}F^{\mu\nu}+\frac{\xi}{2}(A^{\mu}_{\ ;\mu})^2-V,
\label{eqn:abelianlagr}
\end{equation}
where $F_{\mu\nu} = A_{\nu,\mu}-A_{\mu,\nu}$, $V=\lambda(-A_{\mu}A^{\mu})^n$ is the potential with constant $\lambda$ and $n$ belonging to the natural numbers, and  $\xi$ is the gauge-fixing constant.
We assume that the vector field is homogeneus and that the spatial components can be lined up along the $z$ axis, $A_{\mu}= (A_t(t), 0,0,A_z(t))$. In order to formulate a dynamical system, we define the following dimensionless variables: 
\begin{equation}
    x \equiv \frac{\sqrt{\frac{1}{2}\dot{A}_t^2}}{\sqrt{3}m_p H}, \ y \equiv \frac{\sqrt{V}}{\sqrt{3}m_p H},\nonumber
\end{equation}
\begin{equation}
    \ z \equiv \frac{\sqrt{-\frac{1}{2}g^{zz}\dot{A}_z^2}}{\sqrt{3}m_p H},\ w \equiv \sqrt{\frac{3\xi}{2}}\frac{A_t}{m_p}, \ t \equiv \frac{\sqrt{-2 g^{zz}}A_z}{m_p} \,,
    \label{eqn:vector-abelian-free-expansion-var-GF1}
\end{equation}
where $m_p$ is the reduced Planck mass.
The field equations are equivalent to the dynamical system:
\begin{align}
    \Sigma' &= -3 \Sigma +\Xi  t^2 y^2+2 z^2+\Sigma  \epsilon,	
    \label{eqn:U1sigmaeqdyn} \\
    x' &= -\frac{1}{2} \Xi  t^2 w y^2+\frac{2 \Xi  w^3 y^2}{\xi }-\frac{3 w^3}{2}-3 w^2 x+\frac{3 \Sigma ^2 w}{2} \nonumber \\
    & -\frac{3 w x^2}{2} -\frac{2 \Xi  w y^2}{\xi }-\frac{3 w y^2}{2}+\frac{w z^2}{2}+\frac{3 w}{2}+x \epsilon -3 x,
    \label{eqn:U1xeqdyn} \\
    y' &= -\Xi  \Sigma  t^2 y+\frac{1}{2} \Xi  t^2 y-\sqrt{3} \Xi  t y z+\frac{2 \Xi  w x y}{\xi }+y \epsilon, 
    \label{eqn:U1yeqdyn} \\
    w' &= 3 x,
    \label{eqn:U1weqdyn} \\
    z' &= \sqrt{3} \Xi  t y^2-2 \Sigma  z+z \epsilon -2 z,
    \label{eqn:U1zeqdyn} \\
    t' &= \sqrt{2} \Sigma  t-\frac{t}{\sqrt{2}}+\sqrt{6} z,
    \label{eqn:U1teqdyn} \\
   & 1 = \Sigma^2+(x+w)^2 + y^2 +z^2. 
    \label{eqn:abelian-friedmannfree}
\end{align}
The prime denotes differentiation with respect to the e-fold number $N=\ln(a)$, $\epsilon = -\dot{H}/H^2$, and $\Xi=m_p^2 V_{, A^2}/V$.

%------------------------------------------------
%	Critical Points and Stability
%------------------------------------------------

\subsection{Critical Points and Stability for the U(1)-Vector-Field Model}
\label{sec:critabelian}
%\begin{table}[!h]
%\begin{center}
%\begin{tabular}{ c c c c c c c}
%    \hline
%    $\Sigma$& $x$  & $y$   & $w$    & $z$   & $t$ \\
%    \hline
%    0    &  0  &  0  &$\pm1$&  0  &  0 \\
%    $\pm1$ &  0  &  0  &  0   &  0  &  0 \\
%   \hline
%\end{tabular}
%\caption{Critical points and stability.}
%\label{table:abelian-crit}
%\end{center}
%\end{table}
%
%The critical points of the dynamical system are show in the table \ref{table:abelian-crit}.
The first critical point is $\zeta_c =(\Sigma_c,x_c,y_c,w_c,z_c)= (0,0,0,\pm1,0)$ which implies $\epsilon=0$.
Since inflation is equivalent to $\epsilon<1$, this first critical point corresponds to inflation.
%When the parameter $\epsilon<1$, there is inflation.  
%At the first critical point, $\epsilon$ is equal to zero; therefore, this point corresponds to inflation. 
At this point, the energy densities associated with the potential and the Yang-Mills terms are subdominant compared with the energy associated with the gauge-fixing term.
The second critical point is $\zeta_c =(\Sigma_c,x_c,y_c,w_c,z_c)= (\pm1,0,0,0,0)$ which corresponds to an anisotropic state of the system.
To analyze the stability, we translate the critical point to the origin, and then linearize the equations around zero. 
The system of linear differential equations can be expressed as $\mathbf{x}'=\mathbb{M}\mathbf{x}$, where $\mathbb{M}$ is a matrix.
The stability of the system is determined by the sign of the eigenvalues of $\mathbb{M}$. 
If the real part of all the eigenvalues is negative, the critical point is stable. 
If the real part of all the eigenvalues are positive, the critical point is unstable. 
If some eigenvalues have negative real part and the others have positive real part, the critical point is a saddle point and it is unstable. 
When one or more eigenvalues have zero real part, the critical point is nonhyperbolic and to analyze the stability we are requiered to use the centre manifold theorem \cite{wainwright2005dynamical}. 
This theorem guarantees the existence of an invariant manifold which is tangent at the critical point to the space generated by the eigenvectors with  zero-real-part eigenvalues. Analyzing the dynamics of the system on the invariant manifold is easier since this is lower dimensional.

The eigenvalues corresponding to the anisotropic critical point are (6, -3, -3, 3, 3, 3), hence it is unstable.
The eigenvalues corresponding to the inflation critical point are (-3, -3, -3, -2, -1, 0), then it is nonhyperbolic. Using the centre manifold theorem, we found that the dynamics on the invariant manifold is locally governed by the equation:
\begin{equation}
    y'= \Xi y^2/\xi+ \mathcal{O}(y^5).
    \label{eqn:invariantu1}
\end{equation}
For a power-law potential, $\Xi/\xi>0$ when the vector field is timelike. 
Since the temporal part of the field dominates near the critical point, this is unstable. This implies that inflation is a transient period. 

On the other hand, 
%the moment $\epsilon $=1 corresponds to the end of inflation. 
assuming that the system is on the centre manifold, $\epsilon $=1 sets a limit for $y$,
i.e. when the system reaches this value of $y$, inflation ends. 
The system must reach this point after at least 60 efolds in order to solve the classical problems of the standard cosmology.
If the speed of the system is less than 
$y_{\max }/{N}$, inflation will not end before $N$ efolds. 
We obtain the following sufficient condition for inflation to last at least $N$ efolds:
\begin{equation}
    \left|\xi/\lambda\right|\gg 2 n N^{2/3}(-A_0)^{2n-2}/3 H^2|_{\zeta_c}.
    \label{eqn:abelian-temporal-final-result}
\end{equation}

%------------------------------------------------
%	SU(2) Vector Field Model
%------------------------------------------------

\section{SU(2)-Vector-Field Model}
The matter lagrangian density of this model is \cite{Cembranos:2012ng}:
\begin{equation}
        \mathcal{L}_m=-\frac{1}{4} F_{\mu\nu}^a F^{a \mu\nu}  +\frac{\xi}{2}(A_{\ ;\mu}^{\mu\ a})^2 - V( M_{a b}A^{a}_{\mu}A^{b\mu} ),
    \label{eqn:non-abelian-Lmatter}
\end{equation}
where $F_{\mu\nu}^a=A_{\nu,\mu}^a - A_{\mu,\nu}^a + g c^a_{\ b c}A_{\mu}^bA_{\nu}^b$, $c^a_{\ b c}$ are the SU(2)-group-structure constants, $M_{ab}$ is constant $3\times3$ matrix, and $a,b=1,2,3$.
We study the dynamics on the invariant set $A^a_{i}=\dot{A}^a_{i}=0$, with both $A_t^a$ and $\dot{A}_t^a$ being different from zero.
The equations in this particular case can be obtained from the following reduced lagrangian:
\begin{multline}
    L_r = e^{3 \alpha} \left[ -3 m_p^2\left( \dot{\alpha}^2 - \dot{\sigma}^2\right)\right.\\
    + \frac{\xi}{2}(\dot{A}_t^a + 3 A_t^a \dot{\alpha})^2 
\left.- V(M_{a b} A_t^a A_t^b)\right].
    \label{eqn:vector-nonabelian-lagrangian-1azts}
\end{multline}
If $M_{ab}$ is diagonal, the lagrangian \eqref{eqn:vector-nonabelian-lagrangian-1azts} is invariant under ``rotations'' on the $(A^1_t,A^2_t,A^3_t)$ subspace. Therefore, we change the coordinates  on that subspace from cartesian to polar-cylindrical coordinates $r$ and $\theta$.
%\begin{equation}
%   L_r = e^{3\alpha}\left[-3 m_p^2 \left(\dot{\alpha}^2 -\dot{\sigma}^2\right)\right.
%  \left.+\frac{\xi}{2}(\dot{r}^2+r^2\dot{\theta}^2 + 6 \dot{r}r \dot{\alpha} + 9 r^2 \dot{\alpha}^2)-V(r^2)\right].
% \label{eqn:vector-nonabelian-lagrangian-central-prbl}
%\end{equation}
After obtaining the equations from the lagrangian, we express the equations in terms of the following dimensionless variables:
\begin{equation}
    x \equiv \frac{\sqrt{\frac{1}{2}\xi}\dot{r}}{\sqrt{3}\mpl H},\ y \equiv \frac{\sqrt{V}}{\sqrt{3}\mpl H}, 
w \equiv \frac{\sqrt{3\xi}r}{\sqrt{2}\mpl}, 
\ \Theta \equiv \frac{\dot{\theta}}{3H}.
\end{equation}
The field equations are equivalent to the following dynamical system: 
\begin{align}
    \Sigma' &= \Sigma  \left(\epsilon -3\right),\\
    x' &=-\frac{\Theta ^2 w^3}{\xi }-\frac{3 w^3}{2}-3 w^2 x+\sqrt{\frac{3}{2}} \Lambda  w^2 y^2 \nonumber\\
    &+\frac{2 \Theta ^2 w}{\xi }+\frac{3 \Sigma ^2 w}{2}-\frac{3 w x^2}{2}-\frac{3 w y^2}{2}
    +\frac{3 w}{2}\nonumber\\
    &+x \epsilon -3 x-\sqrt{\frac{3}{2}} \Lambda  y^2,\\
    %x' &=-\frac{\Theta ^2 w^3}{\xi }-\frac{3 w^3}{2}-3 w^2 x+\sqrt{\frac{3}{2}} \Lambda  w^2 y^2 \nonumber\\
    %&+\frac{2 \Theta ^2 w}{\xi }+\frac{3 \Sigma ^2 w}{2}-\frac{3 w x^2}{2}-\frac{3 w y^2}{2}+\frac{3 w}{2} \nonumber\\
    %&+x \epsilon -3 x-\sqrt{\frac{3}{2}} \Lambda  y^2,\\
    y' &=y\left(\sqrt{\frac{3}{2}} \Lambda  x + \epsilon \right),\\
    w' &= 3 x,\\
    \Theta' &=\Theta\left( \epsilon-3  -\frac{6  x}{w}\right),
    \label{eqn:dyn} \\
    &1=\Sigma^2+(x+w)^2+y^2+w^2 \Theta^2,
\end{align}
and the parameter $\Lambda=\mpl V_{,r}/\sqrt{\xi}V.$
%        \item The Friedmann equation is: 
%           \begin{equation}
%             \Sigma ^2+\frac{2 \Theta ^2 w^2}{3 \xi }+w^2+2 w x+x^2+y^2 = 1.
%               \label{eqn:non-abelian-fried}
%           \end{equation}
%------------------------------------------------
%	Critical Points and Stability
%------------------------------------------------
\subsection{Inflation Critical Point and Stability}
We find the critical point $\zeta_c =(\Sigma_c,x_c,y_c,w_c,\Theta_c)= (0,0,0,\pm1,0)$.
The parameter $\epsilon$ evaluated at the critical point is equal to 0; therefore, this point corresponds to inflation.
To analyze the stability, we follow the same procedure described in the subsection \ref{sec:critabelian}. 
The eigenvalues associated with this point are (-3, -3, -3, -3, 0), implying that it is nonhyperbolic. 
We are in the need, again, to use the centre manifold theorem. 
The dynamics on the centre manifold follows an equation similar to the one in the abelian case:
\begin{equation}
	y'= \sqrt{3} \Lambda y^3/\sqrt{2} + \mathcal{O}(y^5).
    \label{}
\end{equation}
For a potential of the form $\lambda r^{2n}$, we have $\Lambda>0$, so the point is a saddle point. Finally, analyzing the dynamics on the centre manifold, we obtain the sufficient condition for inflation to last at least $N$ efolds: 
\begin{equation}
    |\sqrt{\xi}/\lambda|\gg \sqrt{6}\mpl n r^{2n-1}N^{2/3}|_{\zeta_c}.
    \label{enqll}
\end{equation}  
\section{Numerical Solution for the SU(2)-Vector-Field Model}
\begin{figure}
\begin{center}
\includegraphics[width=0.3\textheight]{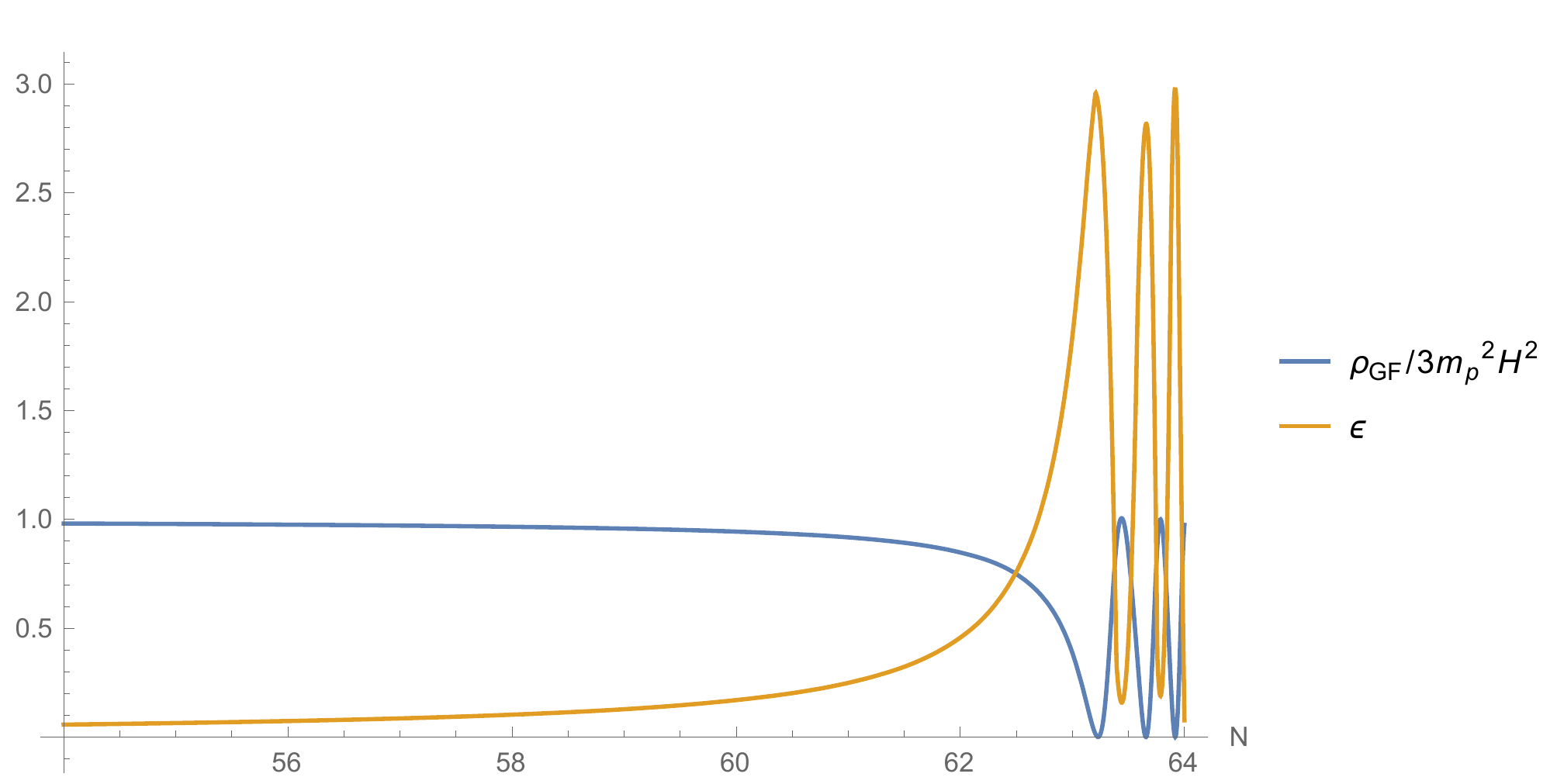}
\caption{ Numerical plots of the normalized energy density of the gauge-fixing term $\rho_{GF}/3m_p^2 H^2$, and $\epsilon$ vs the e-fold number. The plots show that, when  $\rho_{GF}/3m_p^2 H^2$ dominates, there is inflation, and as this term begins to oscillate, inflation ends.}
\label{fig:1}
\end{center}
\end{figure}
%Unfortunately, we can not show all the numerical plots of the solutions of the models. 
The figure \ref{fig:1} shows the numerical solution for the studied non-abelian case. The numerical analysis confirms that the number of efolds depends on the values of $\xi,\lambda,r$ and $H$. The dependence on the gauge-fixing parameter itself is  very weak.
%----------------------------------------------------------------------------------------
%	CONCLUSIONS
%----------------------------------------------------------------------------------------

\section{Conclusions}

The addition of the gauge-fixing term, for the two models studied, implies the existence of an inflationary period which is independent on the choice of the gauge. The only constraint imposed to the choice of the gauge is $\xi>0$, so the fluid does not behave as a phantom.
For both models, the condition for the existence of inflation is $V\ll 3\mpl^2 H^2$. This means that the potential must be negligible compared to $3\mpl^2H^2$ in contrast to scalar-field inflation.
The dynamics of the temporal part of the fields shows that inflation is driven by the gauge-fixing term and lasts long enough to solve the classical problems of cosmology. The number of efolds depends weakly on the choice of the gauge as was expected since the physics must be gauge-independent. 

\vspace{5mm}
{\it Acknowledgements.}
This work was supported by COLCIENCIAS Grant No. 110656933958 RC 0384-2013 and by COLCIENCIAS-ECOS NORD Grant No. RC 0899- 2012 with help of ICETEX.
%----------------------------------------------------------------------------------------
%	FORTHCOMING RESEARCH
%----------------------------------------------------------------------------------------



%% References

%\nocite{*}
\bibliographystyle{elsarticle-num}
\bibliography{referencias}






% \end{thebibliography}

\end{document}